 \documentclass[final,1p,times]{elsarticle}

 \usepackage{graphicx}

\usepackage{amssymb}
\usepackage{amsmath}

\journal{Journal of Computational Physics}

\begin{document}

\begin{frontmatter}



\title{A Fast Numerical Scheme for Causal Relativistic Hydrodynamics with Dissipation}


\author{Makoto Takamoto}

\address{Department of Physics, Kyoto University, Kyoto, 606-8502, Japan}

\author{Shu-ichiro Inutsuka}

\address{Department of Physics, Nagoya University, Nagoya, 464-8602, Japan}

\begin{abstract}
In this paper, 
we develop a stable and fast numerical scheme for relativistic dissipative hydrodynamics 
based on Israel-Stewart theory. 
Israel-Stewart theory is a stable and causal description of dissipation in relativistic hydrodynamics 
although it includes relaxation process with the timescale for collision of constituent particles, 
which introduces stiff equations 
and makes practical numerical calculation difficult. 
In our new scheme, 
we use Strang's splitting method, 
and use the piecewise exact solutions 
for solving the extremely short timescale problem. 
In addition, 
since we split the calculations into inviscid step and dissipative step, 
Riemann solver can be used for obtaining numerical flux for the inviscid step. 
The use of Riemann solver enables us to capture shocks very accurately. 
Simple numerical examples are shown. 
The present scheme can be applied to various high energy phenomena 
of astrophysics and nuclear physics. 
\end{abstract}


\end{frontmatter}


\section{\label{sec:level0}INTRODUCTION}
In recent years, 
various high energy astrophysical phenomena are extensively studied 
by using the relativistic fluid approximation; 
for example, 
ultra relativistic jet~\cite{Begelman et al.(1984),Ghisellini et al.(2005)}, 
GRB~\cite{Sari et al.(1999),Zhang & Meszaros(2004)}, 
Neutron star merger~\cite{Wilson et al.(1996),Shibata et al.(2005)}, 
pulsar wind~\cite{Coroniti(1990),Kirk et al.(2009)}, 
and accretion flows around massive compact objects~\cite{Shakura & Sunyaev(1973),Takeuchi et al.(2010)}. 
In addition, 
because of the recent finding of the strongly coupled 
quark-gluon plasma (sQGP) in the Relativistic Heavy-Ion Collider (RHIC), 
description by relativistic hydrodynamics equations have been vigorously studied 
in the context of nuclear physics~\cite{Baier et al.(2008)}. 
As is well known, 
relativistic fluid equations are highly nonlinear 
because of the Lorentz factor and enthalpy, 
and the high energy astrophysical phenomena are studied mainly by numerical simulation 
except for simplified cases. 
For this reason, 
various numerical formulations of relativistic fluid are investigated. 
However, 
most of the existing numerical schemes assume ideal fluid approximation, 
and there are only few studies 
taking into account the dissipation~\cite{Saida et al.(2010),Bouras et al.(2009),Komissarov 2007,Palenzuela et al.(2009)}. 

Relativistic fluid equation can be obtained 
by tensor decomposition of the particle flux vector and energy-momentum tensor~\cite{deGroot}. 
When one considers ideal fluid, 
those tensors are decomposed 
by assuming homogeneity and isotropy in the comoving frame of each fluid element, 
and dissipation variables are defined as the deviation from the ideal part of those tensors. 
However, 
relativistic fluid includes independent two characteristic directions, 
the particle flux vector and energy flux vector, 
and this results in uncertainty of the definition of the fluid 4-vector. 
For the direction of fluid 4-vector, 
Eckart~\cite{Eckart(1940)} adopted that of particle flux vector, 
and his decomposition is called Eckart formalism; 
Landau and Lifshitz~\cite{Landau & Lifshitz(1959)} adopted that of energy flux, 
and their decomposition is called Landau-Lifshitz formalism. 
In addition to these well-known formalisms, 
various formalisms are proposed~\cite{Tsumura et al.(2007),Tsumura & Kunihiro(2008),Tsumura & Kunihiro(2009)}. 

If one considers the dissipative fluid equations, 
dissipation variables have to be expressed by the fluid variables, density, pressure, and fluid velocity. 
Eckart, Landau, and Lifshitz presented as the expression of dissipation variables 
the relativistic extension of Navier-Stokes approximation. 
However, 
it is well-known that 
the Navier-Stokes equation is parabolic partial differential equations. 
This means that 
the Navier-Stokes equation is acausal, 
and not appropriate for the relativistic equation. 
In addition, 
Hiscock and Lindblom~\cite{Hiscock & Lindblom(1983),Hiscock & Lindblom(1985)} have shown that 
the relativistic Navier-Stokes equations includes unphysical exponentially growing modes, 
and unstable for small perturbation. 
For this problem, 
Israel and Stewart~\cite{Israel & Stewart(1979)} proposed new relativistic dissipation fluid theory 
called Israel-Stewart theory, 
which takes into account the second-order deviation terms for entropy. 
This theory is hyperbolic equations, 
and has been shown to be causal and stable. 
However, 
the equations of Israel-Stewart theory include 14 variables, 
and extremely complex in contrast to relativistic ideal fluid equations. 
In addition, 
the evolution equations of dissipation variables include parameters corresponding to relaxation timescales 
that are much shorter than hydrodynamical timescale, 
and the resultant stiff equations make it difficult to integrate numerically. 
For these reasons, 
applications of Israel-Stewart theory to physical problems are very limited. 

The first numerical scheme of Israel-Stewart theory is proposed by Molnar et al. (2010)~\cite{Molnar et al.(2010)}. 
They integrate fluid equation by using SHASTA as a shock capturing scheme, 
and integrate evolution equations of dissipation variables by using the ordinary explicit difference scheme. 
As explained above, 
evolution equations of dissipation variables are very stiff, 
and they have to use sufficiently high resolution to resolve both the macroscopic and relaxation timescale, 
which demands a exceedingly high numerical cost. 

We develop a new numerical scheme for relativistic dissipative hydrodynamics 
that can integrate accurately and efficiently. 
We split the fluid equations into inviscid part and dissipation part. 
The inviscid part corresponds to ideal relativistic fluid equations, 
and can be solved accurately by using a relativistic Riemann solver~
\cite{Marti & Muller 1994,Marti 1996,Aloy et al. 1999,Pons et al. 2000,Font et al. 2000,
Del Zanna & Bucciantini 2002,Marti & Muller(2003),Mignone & Bodo 2005,Mignone et al.(2005)}. 
The Riemann solver is a method 
to calculate numerical flux by using exact solution of the Riemann problems 
at the interfaces separating numerical grid cells, 
and can be used to describe the flows with strong shocks and sharp discontinuity stably and highly accurately. 
When the dissipation terms are small, 
the dynamics of fluid is dominated by the inviscid part, 
and we can obtain more accurate numerical results by means of the Riemann solver. 
As for the evolution equations of dissipation variables, 
we use the Piecewise Exact Solution method (PES)
~\cite{Inoue et al.(2007),Inoue & Inutsuka(2008),Inoue & Inutsuka(2009)}. 
PES is a numerical method for solving the stiff equation 
by using the formal solution. 
The use of formal solutions for those stiff equations eliminates 
the Courant condition for relaxation timescale. 
In this way, 
our new numerical scheme can describe relativistic dissipative hydrodynamic equations 
highly accurately and efficiently. 
Recently, 
similar methods were applied to solving resistive RMHD~\cite{Komissarov 2007,Palenzuela et al.(2009)}. 

This paper is organized as follows. 
In Sec. \ref{sec:sec1}, 
we present the basic equations of relativistic hydrodynamics. 
We present explicit forms for relaxation timescale parameters 
near the equilibrium. 
In Sec. \ref{sec:sec2}, 
detailed explanation of our new scheme is presented. 
In Sec. \ref{sec:sec3}, 
analyses on stability and causality of relativistic dissipation hydrodynamics equations are presented 
by using simple scalar equation. 
In Sec. \ref{sec:sec4}, 
some results of one-dimensional and multi-dimensional simulations are presented. 

\section{\label{sec:sec1}Basic Equations}
Throughout this paper, we use the units $c = 1$, 
and Cartesian coordinates where the Minkowski metric tensor $\eta_{\mu \nu}$ is given by $\eta_{\mu \nu} = \mathrm{diag}(- 1, 1, 1, 1)$. 
Variables indicated by Greek letters take values from $0$ to $3$, 
and those indicated by Roman letters take values from $1$ to $3$.

We define the convective time derivative $\hat{D}$ 
and the spatial gradient operator $\nabla_{\alpha}$ as follows
\begin{align}
\hat{D} A^{\mu_1 \dots \mu_n} &\equiv u^{\beta} A^{\mu_1 \dots \mu_n}_{;\beta}
,
\\
\nabla_{\alpha} A^{\mu_1 \dots \mu_n} &\equiv \gamma^{\beta}_{\alpha} A^{\mu_1 \dots \mu_n}_{; \beta}
.
\end{align}
where the tensor $\gamma^{\mu \nu}$ is a projection operator on the hyperplane normal to $u^{\mu}$
\begin{align}
\gamma^{\mu \nu} = \eta^{\mu \nu} + u^{\mu} u^{\nu}
\end{align}

%

\subsection{\label{sec:sec1.1}Ideal fluid}
The relativistic hydrodynamic equations can be obtained from the conservation of particle number, momentum, and energy. 
\begin{align}
N^{\mu}_{; \mu} &= 0
\label{eq:d_cons}
,
\\
T^{\mu \nu}_{; \mu} &= 0
\label{eq:em_cons}
,
\end{align}
where $N^{\mu}$ is the particle number density current and 
$T^{\mu \nu}$ the energy-momentum tensor. 
When we consider the ideal fluid, 
they are given by 
\begin{align}
N^{\mu} &= n u^{\mu}
,
\\
T^{\mu \nu} &= \rho h u^{\mu} u^{\nu} + p \eta^{\mu \nu}
,
\end{align}
where $n$ denotes the proper particle number density, 
$p$ is the pressure, 
$h = 1 + \epsilon + p / \rho$ is the specific enthalpy, 
$\epsilon$ is the specific internal energy, 
$\rho \equiv m n$ is the proper rest-mass density, 
and $m$ is the rest-mass of the constituent particle. 
$u^{\mu}$ is the four-velocity of the fluid, 
satisfying the normalization condition: $u^{\mu} u_{\mu} = - 1$. 

When we use Cartesian coordinates, 
the evolution equations of a relativistic fluid are 
\begin{align}
\frac{\partial}{\partial t} 
\left(
 \begin{array}{c}
   D \\
   m^i \\
   E
 \end{array}
\right)
+ \frac{\partial}{\partial x^j} 
\left(
 \begin{array}{c}
   D v^j \\
   m^i v^j + p I^{ij} \\
   m^j
 \end{array}
\right)
= 0,
\end{align}
where ${\bf v}$ is the fluid three-velocity, 
$D, {\bf m}, E$ are the mass, momentum, and energy density relative to the laboratory frame, 
and $I^{ij}$ is the unit tensor. 
In the laboratory frame, $D$, ${\bf m}$, $E$ are given by
\begin{align}
D &= \gamma \rho
, \\
{\bf m} &= \rho h \gamma^2 {\bf v}
, \\
E &= \rho h \gamma^2 - p
,
\end{align}
where $\gamma = (1 - v^2)^{-1/2}$ is the Lorentz factor.
This is the most common form of perfect fluid equations for the numerical hydrodynamics.

\subsection{\label{sec:sec1.2}Causal dissipative fluid}
When one considers relativistic dissipative fluid, 
the system is neither homogeneous nor isotropic, 
and the decomposition of the particle number density current $N^{\mu}$ 
and the energy momentum tensor $T^{\mu \nu}$ 
will change. 
In addition, 
since the dissipative fluid has two different characteristic direction $N^{\mu}$ and $T^{0 \mu}$, 
the definition of the four velocity $u^{\mu}$ is generally not unique, 
that is, 
there is an uncertainty or freedom for our choice of the direction of $u^{\mu}$. 
Since 
one decomposes $N^{\mu}$ and $T^{\mu \nu}$ with respect to $u^{\mu}$, 
this means the form of relativistic dissipative fluid equation is not unique. 
In this paper, 
we consider only Eckart decomposition  
whose four velocity $u^{\mu}$ is parallel to $N^{\mu}$. 
For Landau-Lifshitz decomposition and other decompositions, 
see the following references~\cite{Landau & Lifshitz(1959),Tsumura et al.(2007),Tsumura & Kunihiro(2008),Tsumura & Kunihiro(2009)}. 

In the Eckart decomposition, 
the particle number density current $N^{\mu}$ and the energy-momentum tensor $T^{\mu \nu}$ 
are written as 
\begin{align}
N^{\mu} &= n u^{\mu}
\label{eq:d_N}
,
\\
T^{\mu \nu} &= \rho h u^{\mu} u^{\nu} + p \eta^{\mu \nu} + q^{\mu} u^{\nu} + q^{\nu} u^{\mu} + \tau^{\mu \nu}
\label{eq:d_T}
,
\end{align}
where $q^{\mu}$ is the heat flux vector and $\tau^{\mu \nu}$ is the viscosity tensor. 

From Eqs. (\ref{eq:d_cons}) and (\ref{eq:em_cons}), 
the evolution equations of relativistic dissipative fluid are given by 
\begin{align}
\frac{\partial}{\partial t} 
\left(
 \begin{array}{c}
   D \\
   m^i + q^0 u^i + q^i u^0 + \tau^{0 i} \\
   E + 2 q^0 u^0 + \tau^{00}
 \end{array}
\right)
+ \frac{\partial}{\partial x^j}
\left(
 \begin{array}{c}
   D v^j \\
   m^i v^j + p I^{i j} + q^i u^j + q^j u^i + \tau^{i j} \\
   m^j + q^0 u^j + q^j u^0 + \tau^{0 j} \\
 \end{array}
\right)
= 0,
\label{eq:d_eqs}
\end{align}
In contrast to the non-relativistic case, 
dissipation variables are differentiated with respect to time. 
For example, 
time derivative of the heat flux vector remains in the second line of Eq. (\ref{eq:d_eqs}) 
even in the fluid rest frame. 
This is because 
the energy flux is identified with the momentum density. 
Then, 
if one uses the relativistic Navier-Stokes terms as the dissipative ones, 
the fluid equations becomes parabolic, 
and the characteristic velocity of this theory becomes infinity. 
This means that 
the dissipation variables evolve into equilibrium values within infinitely short time, 
and the time derivative of them diverges. 
Hiscock and Lindblom~\cite{Hiscock & Lindblom(1983),Hiscock & Lindblom(1985)} proved that 
the relativistic Navier-Stokes theory adopting any definitions of four-velocity is unstable
in the sense that small perturbation will diverge exponentially with time 
in any frame 
except for Landau-Lifshitz theory in its rest frame. 
In order to find stable and causal theory, 
Israel and Stewart developed the following second-order theory 
from relativistic Boltzmann equation 
%
\begin{align}
\hat{D} \Pi &= \frac{1}{\tau_{\Pi}} (\Pi_{NS} - \Pi) -I_{\Pi}
\label{eq:bulk_IS}
,
\\
\hat{D} \pi^{\mu \nu} &= \frac{1}{\tau_{\pi}} (\pi^{\mu \nu}_{NS} - \pi^{\mu \nu}) - I_{\pi}^{\mu \nu}
\label{eq:vis_IS}
,
\\
\hat{D} q^{\mu} &= \frac{1}{\tau_{q}} (q^{\mu}_{NS} - q^{\mu}) - I_q^{\mu}
\label{eq:heat_IS}
,
\end{align}
where $\pi^{\mu \nu}$ and $\Pi$ are the shear viscosity and bulk viscosity 
defined as the traceless part and trace part of the viscosity tensor $\tau^{\mu \nu}$ respectively, 
\begin{equation}
  \tau^{\mu \nu} \equiv \Pi \gamma^{\mu \nu} + \pi^{\mu \nu}
  ,
\end{equation}
$\tau_q$, $\tau_{\Pi}$, and $\tau_{\pi}$ are the relaxation times, 
and $I$ is second-order terms, 
which are the product of dissipation variables and derivative of fluid variables.  
Note that 
they can be neglected in the astrophysical application,  
since the gradient of fluid variables are not so steep. 
If one considers the application to the QGP, 
one can use the following abbreviations~\cite{Molnar et al.(2010)}
\begin{align}
  I_{\Pi} &= \frac{1}{2} \Pi \left( \nabla_{\lambda} u^{\lambda} + D \ln \frac{\beta_0}{T} \right)
  \label{eq:I_pi}
  , 
  \\
  I^{\mu \nu}_{\pi} &= (\pi^{\lambda \mu} u^{\nu} + \pi^{\lambda \nu} u^{\mu}) D u_{\lambda} 
  \\
  &+ \frac{1}{2} \pi^{\mu \nu} \left( \nabla_{\lambda} u^{\lambda} + D \ln \frac{\beta_2}{T} \right) 
  \\
  &+ \pi^{\mu \lambda} \omega^{\nu}_{\lambda} + \pi^{\nu \lambda} \omega^{\mu}_{\lambda}
  \label{eq:I_tau}
  , 
\end{align}
where $\omega$ is the rotational part of $u^{\mu}_{; \nu}$ defined as 
\begin{equation}
  \omega^{\mu}_{\nu} \equiv \frac{1}{2} \gamma^{\mu \alpha} \gamma^{\beta}_{\nu} (u_{\alpha ; \beta} - u_{\beta ; \alpha})
  .
\end{equation}
When one adopts Landau-Lifshitz frame as is often the case with QGP applications, 
$I_q$ vanishes~\cite{Molnar et al.(2010)}. 

The ``second-order'' means that 
the entropy current contains second-order terms 
in deviations from equilibrium~\cite{israel1963,Israel & Stewart(1979)}. 
The above equations take into account time derivative of dissipation variables, 
that is, relaxation effect. 
Owing to these terms, 
the relativistic dissipative fluid equations become hyperbolic, 
and it allows the equations become stable and causal 
if one uses appropriate parameters. 
We discuss these appropriate parameters in Sec \ref{sec:sec3}. 

The Navier-Stokes terms are given by 
\begin{align}
q_{NS}^{\mu} &= - \kappa \gamma^{\mu \nu} \left( T_{,\nu} + T u^{\rho} u_{\nu, \rho} \right)
,
\label{eq:th_cu_ext}
\\
\tau_{NS}^{\mu \nu} &= \pi^{\mu \nu}_{NS} + \Pi_{NS} \gamma^{\mu \nu}
\nonumber
\\
&= - \gamma^{\mu \rho} \gamma^{\nu \sigma} 
\left[ \eta \left(u_{\rho, \sigma} + u_{\sigma, \rho} - \frac{2}{3} \eta_{\rho \sigma}
u^{\lambda}_{,\lambda} \right) \right] - \zeta u^{\lambda}_{, \lambda} \gamma^{\mu \nu} 
,
\label{eq:vis_tens}
\end{align}
where $\kappa$ is the heat conduction coefficient, 
$\eta$ is the shear viscosity coefficient, 
and $\zeta$ is the bulk viscosity coefficient. 
The subscript $NS$ means ``Navier-Stokes'' terms. 
If dissipation terms are small, 
we can rewrite Eq. (\ref{eq:th_cu_ext}) by substituting ideal equation of motion into $u^{\rho} u_{\nu, \rho}$, 
and obtain 
\begin{align}
q_{NS}^{\mu} &= - \kappa \gamma^{\mu \nu} \left( T_{,\nu} - \frac{T}{\rho h} p_{,\nu} \right)
. 
\label{eq:th_cu}
\end{align}
In this paper, 
we use Eq. (\ref{eq:th_cu}) as the Navier-Stokes heat flux vector. 

Heat flux vector has $4$ components, 
and viscosity tensor $10$ components. 
However, 
these are constrained by the following orthogonality conditions: 
\begin{align}
\tau^{i 0} u_0 &= - \tau^{i j} u_j
,
\label{eq:orth1}
\\
\tau^{00} u_0 &= - \tau^{0 j} u_j
,
\label{eq:orth2}
\\
q^0 u_0 &= - q^j u_j
.
\label{eq:orth3}
\end{align}
As a result, 
the number of physical degrees of freedom reduces to $3$ and $6$ respectively. 

Eqs. (\ref{eq:vis_tens}) and (\ref{eq:th_cu}) are given in the covariant form. 
When one considers the flat Cartesian coordinate and one-dimensional problem, 
they reduce to 
\begin{align}
\tau_{NS}^{0 x} &= - \gamma^{0 \rho} \gamma^{x \sigma} 
\left[\eta( \partial_{\rho} u_{\sigma} + \partial_{\sigma} u_{\rho} )
+ \left( \zeta - \frac{2}{3} \eta \right) g_{\rho \sigma} \partial_{\lambda} u^{\lambda} \right]
\nonumber
\\
&= - \left[
\eta \left\{
u^x u^t \partial_t u^t + (- 1 + (u^t)^2 ) \partial_t u^x + (1 + (u^x)^2 ) \partial_x u^t + u^t u^x \partial_x u^x
\right\}
\right.
\nonumber
\\
&+\left. \left( \zeta - \frac{2}{3} \eta \right) u^0 u^x \theta
\right]
\label{eq:NS_1st}
,
\\
\tau_{NS}^{0 \perp} &= - \gamma^{0 \rho} \gamma^{\perp \sigma} 
\left[\eta( \partial_{\rho} u_{\sigma} + \partial_{\sigma} u_{\rho} )
+ \left( \zeta - \frac{2}{3} \eta \right) g_{\rho \sigma} \partial_{\lambda} u^{\lambda} \right]
\nonumber
\\
&= - \left[
\eta \left\{
u^{\perp} u^t \partial_t u^t + (- 1 + (u^t)^2 ) \partial_t u^{\perp} + u^{\perp} u^x \partial_x u^t + u^t u^x \partial_x u^{\perp}
\right\}
\right.
\nonumber
\\
&+\left. \left( \zeta - \frac{2}{3} \eta \right) u^0 u^{\perp} \theta
\right]
,
\end{align}
\begin{align}
\tau_{NS}^{y z} &= - \gamma^{y \rho} \gamma^{z \sigma} 
\left[\eta( \partial_{\rho} u_{\sigma} + \partial_{\sigma} u_{\rho} )
+ \left( \zeta - \frac{2}{3} \eta \right) g_{\rho \sigma} \partial_{\lambda} u^{\lambda} \right]
\nonumber
\\
&= - \left[
\eta \left\{
u^y u^t \partial_t u^z + u^z u^t \partial_t u^y + u^y u^x \partial_x u^z + u^z u^x \partial_x u^y
\right\}
\right.
\nonumber
\\
&+\left. \left( \zeta - \frac{2}{3} \eta \right) u^y u^z \theta
\right]
,
\\
\tau_{NS}^{x \perp} &= - \gamma^{x \rho} \gamma^{\perp \sigma} 
\left[\eta( \partial_{\rho} u_{\sigma} + \partial_{\sigma} u_{\rho} )
+ \left( \zeta - \frac{2}{3} \eta \right) g_{\rho \sigma} \partial_{\lambda} u^{\lambda} \right]
\nonumber
\\
&= - \left[
\eta \left\{
u^{\perp} u^t \partial_t u^x + u^x u^t \partial_t u^{\perp}  + u^{\perp} u^x \partial_x u^x + (1 + (u^x)^2 ) \partial_x u^{\perp}
\right\}
\right.
\nonumber
\\
&+\left. \left( \zeta - \frac{2}{3} \eta \right) u^x u^{\perp} \theta
\right]
,
\\
\tau_{NS}^{x x} &= - \gamma^{x \rho} \gamma^{x \sigma} 
\left[\eta( \partial_{\rho} u_{\sigma} + \partial_{\sigma} u_{\rho} )
+ \left( \zeta - \frac{2}{3} \eta \right) g_{\rho \sigma} \partial_{\lambda} u^{\lambda} \right]
\nonumber
\\
&= - \eta \partial_x u^x - \left( \zeta - \frac{2}{3} \eta \right) (1 + (u^x)^2) \theta
,
\\
q_{NS}^x &= - \gamma^{x \mu} \kappa \left[ \partial_{\mu} - \frac{T}{\rho h} \partial_{\mu} p \right]
\nonumber
\\
&= - \kappa \left[
u^x u^t \left(\partial_t T - \frac{T}{\rho h} \partial_t p \right) 
+ (1 + (u^x)^2) \left(\partial_x T - \frac{T}{\rho h} \partial_x p \right)
\right]
,
\\
q_{NS}^{\perp} &= - \gamma^{\perp \mu} \kappa \left[ \partial_{\mu} - \frac{T}{\rho h} \partial_{\mu} p \right]
\nonumber
\\
&= - \kappa u^{\perp} \left[
u^t \left(\partial_t T - \frac{T}{\rho h} \partial_t p \right) 
+ u^x \left(\partial_x T - \frac{T}{\rho h} \partial_x p \right)
\right]
\label{eq:NS_last}
,
\end{align}
where $\theta$ is the expansion of the fluid defined as 
\begin{align}
\theta \equiv \nabla_{\mu} u^{\mu} 
= \partial_{\mu} u^{\mu} + \Gamma^{\mu}_{\alpha \mu} u^{\alpha}
.
\end{align}

In the above expressions, 
the viscosity tensor includes both shear viscosity and bulk viscosity. 
This is because 
this form is useful for the directional splitting explained in Sec. \ref{sec:sec2.4}. 

\subsection{\label{sec:sec1.3}Explicit forms of relaxation time}
Israel-Stewart theory includes some transport coefficients, 
that is, dissipation coefficients and relaxation times, 
and their values depend on distribution functions 
and the cross sections of collisions between constituent particles. 
Since we are interested in the fluid approximation, 
we consider only the cases close to the equilibrium~\cite{Israel & Stewart(1979),deGroot}. 

In this case, 
the explicit forms of relaxation time can be expressed as follows:
\begin{equation}
  \tau_{\Pi} = \frac{\zeta \beta_0}{3}
  , 
  \quad 
  \tau_q = \kappa T \beta_1
  ,
  \quad
  \tau_{\pi} = 2 \eta \beta_2
\end{equation}
where
\begin{align}
  \Gamma / (\Gamma - 1) &= \beta^2 (1 + 5 h / \beta - h^2)
  , 
  \\
  \alpha_0 &= (\Gamma - 1) \Omega^{**} / \Gamma \Omega p, \quad \alpha_1 = - (\Gamma - 1) \Gamma p
  ,
  \\
  \beta_0 &= \frac{3 \Omega^{*}}{h^2 \Omega^2 p}
  , 
  \qquad 
  \beta_1 = \left( \frac{\Gamma - 1}{\Gamma} \right)^2 \frac{\beta}{h p} \left( 5 h^2 - \frac{\Gamma}{\Gamma - 1} \right)
  , 
  \\
  \beta_2 &= \frac{1 + 6 h / \beta}{2 h^2 p}
  ,
  \qquad
  a_1 = - \frac{1 + h \beta (\Gamma - 1) / \Gamma }{h^2 p}
  ,
  \\
  \Omega &= 3 \Gamma - 5 + 3 \Gamma / h \beta
  ,
  \quad
  \Omega^* = 5 - 3 \Gamma + 3 (10 - 7 \Gamma) h / \beta
  , 
  \\
  \Omega^{**} &= 5 - 3 \Gamma + 3 \Gamma^2 / (\Gamma - 1) h^2 \beta^2
  ,
\end{align}
$\beta = m / T$, 
and $h$ is the enthalpy. 

Asymptotic forms of relaxation time are:
\begin{enumerate}
  \item In the non-relativistic limit ($\beta \rightarrow \infty, \quad \Gamma = 5 / 3$)
  \begin{equation}
    \tau_{\Pi} = \frac{2}{5} \frac{\zeta \beta^2}{p}
    , 
    \quad
    \tau_{q} = \frac{2 \kappa T \beta}{5 p}
    , 
    \quad
    \tau_{\pi} = \frac{\eta}{p}
    \label{eq:relax_nonrela}
  \end{equation}
  
  \item In the ultra-relativistic limit ($\beta \rightarrow 0, \quad \Gamma = 4 / 3$)
  \begin{equation}
    \tau_{\Pi} = \frac{72 \zeta}{\beta^4 p}
    ,
    \quad
    \tau_{q} = \frac{5 \kappa T}{4 p}
    ,
    \quad
    \tau_{\pi} = \frac{3 \eta}{2 p}
    \label{eq:relax_ultrarela}
  \end{equation}
\end{enumerate}

The explicit forms of dissipation coefficients $\kappa, \eta, \zeta$ 
depend on the particle interaction, 
and one can use appropriate dissipation coefficients for each problems. 

\section{\label{sec:sec2}Numerical scheme}
In this section, 
we start description of our scheme for one-dimensional case. 
Multi-dimensional case is shown in Sec \ref{sec:sec2.4}.
\subsection{\label{sec:sec2.1}Strang splitting method}
The relativistic dissipative fluid equation is the hyperbolic-relaxation one, 
and this has different difficulty for the inviscid and dissipation part respectively. 
The difficulty of inviscid part of fluid equation results from the non-linearity of the fluid equation, 
and this exists even in non-relativistic ideal fluid equation; 
the difficulty of dissipation part is that 
the evolution equations of dissipation Eqs. (\ref{eq:bulk_IS}), (\ref{eq:vis_IS}), and (\ref{eq:heat_IS}) 
are stiff equations. 
When one solves stiff equation by using explicit differentiation, 
$\Delta t$ must be shorter than relaxation timescale parameters for stability. 
However, relaxation timescale parameters $\tau_{\Pi}$, $\tau_{\pi}$, $\tau_q$ are generally much shorter than 
dynamical timescale of fluid, 
and it needs a heavy computational cost. 

In order to address these problems separately, 
we apply Strang splitting method 
and split the relativistic dissipative fluid equation as follows 
\begin{align}
\frac{\partial}{\partial t} 
\left(
 \begin{array}{c}
   D \\
   m^i \\
   E 
 \end{array}
\right)
+ \frac{\partial}{\partial x^j}
\left(
 \begin{array}{c}
   D v^j \\
   m^i v^j + p I^{i j} \\
   m^j \\
 \end{array}
\right)
= 0,
\label{eq:ideal}
\end{align}
\begin{align}
\frac{\partial}{\partial t} 
\left(
 \begin{array}{c}
   D \\
   m^i + q^0 u^i + q^i u^0 + \tau^{0 i} \\
   E + 2 q^0 u^0 + \tau^{00}
 \end{array}
\right)
+ \frac{\partial}{\partial x^j}
\left(
 \begin{array}{c}
   0 \\
   q^i u^j + q^j u^i + \tau^{i j} \\
   q^0 u^j + q^j u^0 + \tau^{0 j} \\
 \end{array}
\right)
= 0,
\label{eq:dissip}
\end{align}

First, 
the inviscid part Eq. (\ref{eq:ideal}) can be solved accurately 
by using the Riemann solver. 
The Riemann solver is a method 
that calculates numerical flux by using exact or approximate solution of the Riemann problem at the cell boundary, 
and it is known that 
this method is stable and accurate. 

Next, 
we consider the dissipation part Eq. (\ref{eq:dissip}). 
The second terms of Eq. (\ref{eq:dissip}) are the remaining part of the flux of Eq. (\ref{eq:d_eqs}), 
and includes dissipation variables and four velocity $u^{\mu}$. 
For the second-order accuracy in time, 
one must use states evolved half time-step $\Delta t / 2$. 
The first terms of Eq. (\ref{eq:dissip}) are conserved variables, 
and they includes dissipation variables unlike non-relativistic case. 
In order to calculate stably, 
one has to substitute for these dissipation variables 
not the Navier-Stokes terms, 
but evolved ones by Eqs. (\ref{eq:bulk_IS}), (\ref{eq:vis_IS}), (\ref{eq:heat_IS}). 
Then, 
the first terms of Eq. (\ref{eq:dissip}) includes four velocity $u^{\mu}$. 
For this reason, 
one has to calculate them before the calculation of inviscid part, 
and save them until the dissipation part. 

In summary, 
we split the conserved variables $U$ of Eq. (\ref{eq:dissip}) as follows
\begin{equation}
U = U_{ideal} + U_{dissip}
,
\end{equation}
where
\begin{align}
U_{ideal} = 
\left(
 \begin{array}{c}
   D \\
   m^i \\
   E 
 \end{array}
\right)
,
\quad
U_{dissip} = 
\left(
 \begin{array}{c}
   0 \\
   q^0 u^i + q^i u^0 + \tau^{0 i} \\
   2 q^0 u^0 + \tau^{00}
 \end{array}
\right)
.
\end{align}
First, 
we calculate $U = U_{ideal} + U_{dissip}$. 
Then, 
we evolve $U_{ideal}$ by using Riemann solver. 
Next, 
we calculate $\tilde{U}_{ideal} + U_{dissip}$ as the initial value of $U$ of dissipation step, 
and integrate Eq. (\ref{eq:dissip}) over the full time-step. 
$\tilde{U}_{ideal}$ is the ideal part of conserved variable evolved by using Riemann solver. 
By using the conservation-law form for Eqs. (\ref{eq:ideal}) and (\ref{eq:dissip}), 
this method satisfies the conservation law of mass, momentum, and energy 
within machine round-off error. 


In addition, 
we adopt $q^x, q^y, q^z, \tau^{0x}, \tau^{0y}, \tau^{0z}, \tau^{yz}, \tau^{zx}, \tau^{xy}$ 
for the primitive variables of the dissipation. 
This selection is based on the equality of the spatial direction, 
and is useful for the directional split. 
The other variables can be calculated 
  by the orthogonality conditions Eqs. (\ref{eq:orth1}), (\ref{eq:orth2}), and (\ref{eq:orth3}). 
Since the dissipation variables are necessary for the primitive recovery procedure, 
we define the variables at the cell-center. 
For the calculation of the numerical flux at the cell-boundary, 
we use the average of the dissipation variables, 
and evolve timestep $\Delta t / 2$ at the cell-boundary. 

\subsection{\label{sec:sec2.2}Stiff equation}
When one describes relativistic dissipative fluid by using the Israel-Stewart theory, 
one has to take into account the evolution of dissipation variables. 
However, 
those equations are stiff-equations, 
that is, 
the form of those equations are given by 
\begin{equation}
\partial_t U = \frac{S(U)}{\tau_{relax}}
\label{eq:stiff}
,
\end{equation}
where $\tau_{relax}$ is the relaxation time, 
and this is the characteristic timescale of the evolution of $U$. 
If this timescale $\tau$ is much shorter than the fluid timescale $\tau_{fluid}$, 
Eq. (\ref{eq:stiff}) is called ``stiff-equation'', 
and the stability of an explicit scheme is achieved only 
with a timestep size $\Delta t \lesssim \tau_{relax} \ll \tau_{fluid}$. 
In general, this constraint is more restrictive than the Courant-Friedrichs-Lewy (CFL) condition 
$\Delta t \le \Delta x / c_{charact}$, 
where $c_{charact}$ is the characteristic velocity of fluid, 
for example, sound velocity, Alfv\'en velocity, and so on. 
This increases the computational cost exceedingly, 
which hinders the use of simple explicit scheme. 
For avoiding this timestep restriction, 
we apply the Strang-splitting technique. 

First, 
rewriting Eqs. (\ref{eq:bulk_IS}), (\ref{eq:vis_IS}), (\ref{eq:heat_IS}) 
in the coordinate dependent forms, 
they reduce to 
\begin{align}
\gamma \left(\frac{\partial}{\partial t} + v^j \frac{\partial}{\partial x^j}\right) \Pi 
&= \frac{1}{\tau_{\Pi}} (\Pi_{NS} - \Pi) -I_{\Pi}
,
\\
\gamma \left(\frac{\partial}{\partial t} + v^j \frac{\partial}{\partial x^j}\right) \pi^{\mu \nu} 
&= \frac{1}{\tau_{\pi}} (\pi^{\mu \nu}_{NS} - \pi^{\mu \nu}) - I_{\pi}
,
\\
\gamma \left(\frac{\partial}{\partial t} + v^j \frac{\partial}{\partial x^j}\right) q^{\mu} 
&= \frac{1}{\tau_{q}} (q^{\mu}_{NS} - q^{\mu}) - I_q
. 
\end{align}
Then, 
we split the above equations as follows 
\begin{align}
\left(\frac{\partial}{\partial t} + v^j \frac{\partial}{\partial x^j}\right) \Pi 
&=  - \frac{I_{\Pi}}{\gamma}
\label{eq:adv_bulk}
,
\\
\left(\frac{\partial}{\partial t} + v^j \frac{\partial}{\partial x^j}\right) \pi^{\mu \nu} 
&= - \frac{I_{\pi}}{\gamma}
\label{eq:adv_vis}
,
\\
\left(\frac{\partial}{\partial t} + v^j \frac{\partial}{\partial x^j}\right) q^{\mu} 
&= - \frac{I_q}{\gamma}
\label{eq:adv_heat}
,
\end{align}
and 
\begin{align}
\frac{\partial}{\partial t} \Pi 
&= \frac{1}{\gamma \tau_{\Pi}} (\Pi_{NS} - \Pi)
\label{eq:relax_bulk}
,
\\
\frac{\partial}{\partial t} \pi^{\mu \nu} 
&= \frac{1}{\gamma \tau_{\pi}} (\pi^{\mu \nu}_{NS} - \pi^{\mu \nu})
\label{eq:relax_vis}
,
\\
\frac{\partial}{\partial t} q^{\mu} 
&= \frac{1}{\gamma \tau_{q}} (q^{\mu}_{NS} - q^{\mu})
\label{eq:relax_heat}
. 
\end{align}
Eqs. (\ref{eq:adv_bulk}), (\ref{eq:adv_vis}), and (\ref{eq:adv_heat}) 
are the first-order advection equation with source terms, 
and one can solve them by upwind scheme. 
Eqs. (\ref{eq:relax_bulk}), (\ref{eq:relax_vis}), and (\ref{eq:relax_heat}) 
are stiff equation 
that requires special care. 
In our new method, 
we solve the stiff equations by using the PES method
~\cite{Inoue et al.(2007),Inoue & Inutsuka(2008),Inoue & Inutsuka(2009)}. 
Since we use Strang-splitting technique, 
one can obtain the formal solutions of 
Eqs. (\ref{eq:relax_bulk}), (\ref{eq:relax_vis}), and (\ref{eq:relax_heat}) as follows: 
\begin{align}
\Pi &= (\Pi_0 - \Pi_{NS}) \exp \left[- \frac{t - t_0}{\tau_{\Pi}} \right] + \Pi_{NS}
\label{eq:PES_b}
,
\\
\pi^{\mu \nu} &= (\pi^{\mu \nu}_0 - \pi^{\mu \nu}_{NS}) \exp \left[- \frac{t - t_0}{\tau_{\pi}} \right] + \pi^{\mu \nu}_{NS}
\label{eq:PES_s}
,
\\
q^{\mu} &= (q^{\mu}_0 - q^{\mu}_{NS}) \exp \left[- \frac{t - t_0}{\tau_q} \right] + q^{\mu}_{NS}
\label{eq:PES_q}
,
\end{align}
where subscript $0$ means the initial value. 
Since these are the formal solution, 
numerical calculation of these terms remains stable irrespective of the timestep. 
In this way, 
the time-step of our scheme is not restricted by the stiff equations 
Eqs. (\ref{eq:relax_bulk}), (\ref{eq:relax_vis}), and (\ref{eq:relax_heat})
those correspond to relaxation. 
In this way, 
we solve the stiff equation by using piecewise exact solution Eqs. (\ref{eq:PES_b}) - (\ref{eq:PES_q}). 
Thus, this procedure is called PES method. 
Note that the terms $I$; 
$I$ in Eqs. (\ref{eq:I_pi}) and (\ref{eq:I_tau}) include dissipation variables, 
which might complicate the actual exact solution. 
In this paper, 
we recommend to assume that 
the dissipation variables in these terms are constant, 
and simply add them to the Eqs. (\ref{eq:PES_b}) - (\ref{eq:PES_q}). 
If this procedure results in bad approximation, 
it means that 
one should not use fluid approximation 
but the kinetic equation. 
This is because 
these terms should be small compare to the other dissipation terms 
when fluid approximation is justified. 

The Navier-Stokes terms Eqs. (\ref{eq:NS_1st}) - (\ref{eq:NS_last}) include 
not only spatial derivatives 
but also time derivatives unlike non-relativistic case. 
We calculate time derivatives by using the following form of first-order explicit finite differentiation 
\begin{equation}
\partial_t U^n = \frac{\hat{U}^{n+1}_{ideal} - U^n}{\Delta t}
. 
\end{equation}
where $\hat{U}^{n+1}_{ideal}$ is the variable evolved by the inviscid step, 
and $U^n$ is the initial value of the n-th step. 
For this reason, 
one has to save initial fluid variables until dissipation step. 
In addition, 
we approximate the spatial derivatives of the Navier-Stokes terms with the centered finite differences. 
This is because 
the physical meanings of the dissipation variables are the diffusion. 
For other part of the spatial derivatives, 
we use the MUSCL scheme by Van Leer~\cite{VanLeer} for the second-order accuracy in space. 

\subsection{\label{sec:sec2.3}Primitive recovery}
When one uses conservation-law form, 
updated variables are not primitive variables but conserved variables, 
and one has to obtain the primitive variables from conserved variables. 
In the case of relativistic fluid, 
one has to solve a non-linear algebraic equation for this primitive recovery 
even in the ideal fluid case. 
In the case of relativistic dissipative fluid, 
the primitive recovery is more complicated, 
since the conserved variables include dissipation terms. 
In this section, 
we explain a method that makes the primitive recovery somewhat simple. 

The dissipation variables in the conserved ones are obtained 
by using relaxation equations Eqs. (\ref{eq:bulk_IS}), (\ref{eq:vis_IS}), and (\ref{eq:heat_IS}). 
Then, 
a major cause of the complexity of the primitive recovery is 
four velocity $u^{\mu}$ multiplied by dissipative variables. 
If the effect of the dissipation is small enough, 
one can expect that 
the change of four velocity $u^{\mu}$ during dissipation step is smaller than during inviscid step. 
For this reason, 
we use four velocity $u^{\mu}$ obtained after inviscid step as the initial guess, 
and this enable us to calculate the dissipation part of conserved variables $U_{dissip}$. 
We can obtain ideal part $U_{ideal}$ by subtracting $U_{dissip}$ from conserved variables $U$, 
and we can obtain primitive variables by using the primitive recovery for ideal fluid part. 
Then, 
we replace the initial guess for $u^{\mu}$ in $U_{dissip}$ with the obtained four velocity, 
and carry out the primitive recovery again. 
We repeat this procedure until the primitive variables converge. 
In this way, 
one can obtain primitive variables consistently. 

In summary, 
\begin{enumerate}
\item 
Calculate $U_{dissip}$ using $u^{\mu}$ obtained by the inviscid step as an initial guess, 
and evolve dissipative variables. 
\item 
Calculate $U_{ideal} = U - U_{dissip}$, 
and obtain primitive variables by using ordinary primitive recovery of relativistic ideal fluid. 
For the stability of the numerical calculation, 
we have imposed the stability condition $|U| > c |U_{dissip}|$, 
where $c$ is some constant number smaller than unity. 
\item
Replace the initial guess for $u^{\mu}$ in $U_{dissip}$ with the obtained four velocity, 
and carry out the primitive recovery again. 
\item
Repeat this procedure until the primitive variables converge. 
\end{enumerate}

From test calculations, 
this approach seems to work well even for large dissipation coefficients 
and discontinuous profiles for physical variables, 
and converges within less than 5 iterations. 

\subsection{\label{sec:sec2.4}multi-dimensional case}
We have explained one-dimensional scheme so far. 
For multidimensional calculation, 
we can apply the directional splitting method~\cite{strang} 
where one applies one-dimensional operator in each spatial direction successively. 
To achieve second-order accuracy in time, 
one has to apply one-dimensional operator in the following order 
for two-dimensional case 
\begin{equation}
U^{n+1} = L_x^{1/2} L_y L_x^{1/2} U^{n}
,
\end{equation}
and for three-dimensional case 
\begin{equation}
U^{n+1} = L_x^{1/6} L_y^{1/6} L_z^{1/3} 
L_y^{1/6} L_x^{1/3} L_z^{1/6} 
L_y^{1/3} L_x^{1/6} L_z^{1/3}
L_x^{1/6} L_y^{1/3} L_z^{1/6} 
L_x^{1/6} U^{n}
.
\end{equation}
The $\Delta t$ can be determined by Courant-Friederichs-Lewy (CFL) condition 
presented in the next section. 


\section{\label{sec:sec3}Causality and Stability}
The Israel-Stewart theory is known as the stable and causal relativistic dissipative fluid theory. 
Strictly speaking, 
one has to use the appropriate parameters for the stability and causality. 
However, 
in order to guarantee the stability and causality, 
the values of the various dissipation coefficients should be limited to certain ranges. 
In this section, 
we discuss these parameters. 

\subsection{\label{sec:sec3.1}Stability of the telegrapher equation}
The essential structure of the Israel-Stewart theory can be analyzed by the following simple equations 
\begin{align}
\partial_t Q &+ \nabla \cdot {\bf F} = 0
\label{eq:advec_Q}
,
\\
\partial_t {\bf F} &= - \frac{1}{\tau} ({\bf F} + \eta \nabla Q)
\label{eq:relax_F}
.
\end{align}
Eliminating ${\bf F}$ from the above equations, 
one obtains the following telegrapher equation 
\begin{equation}
\partial_t^2 Q + \frac{1}{\tau} \partial_t Q - \frac{\eta}{\tau} \triangle Q = 0
\label{eq:teleg}
. 
\end{equation}
Then, 
the characteristic velocity of Eq. (\ref{eq:teleg}) 
is given by 
\begin{equation}
v_c = \sqrt{\frac{\eta}{\tau}}
\end{equation}
For the causality, 
the characteristic velocity must be lower than the maximum physical velocity. 
Thus, 
the relaxation timescale parameter $\tau$ has the following physical lower limit 
\begin{equation}
\tau \ge \tau_{min} \equiv \frac{\eta}{v^2_{\mathrm{max}}}
\label{eq:causal}
,
\end{equation}
where $v_{{\mathrm{max}}}$ is the maximum velocity of the physical system, 
that is smaller than the speed of light. 

Next, 
we consider the stability condition. 
Israel and Stewart~\cite{Israel & Stewart(1979)} indicates that 
the Israel-Stewart theory is stable and causal theory 
if one considers the Boltzmann gas, 
that is, near the equilibrium distribution. 
When one uses a simple explicit finite difference scheme, 
the CFL condition is that 
the timestep size $\Delta t$ is less than the relaxation time $\tau$: $\Delta t < \tau$. 
In the following, 
we consider the stability condition of PES method for Eqs. (\ref{eq:advec_Q}) and (\ref{eq:relax_F}) 
by using the von Neumann's stability analysis 
and numerical simulation. 

\begin{figure}[here]
 \centering
  \includegraphics[width=7cm]{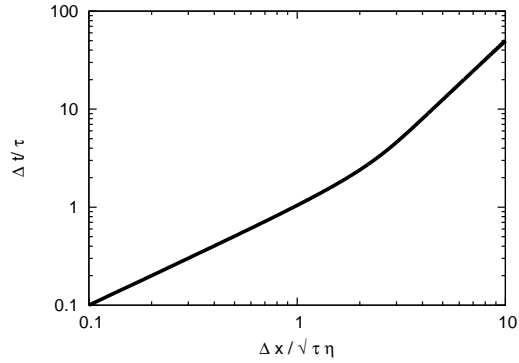}
  \caption{Numerical solution of Eq. (\ref{eq:vonNeumann_PES}) when $|R| = 1$. 
           This figure shows that 
           $\Delta t /\tau$ is proportional to $\Delta x$ when $\Delta t / \tau \ll 1$, 
           and $\Delta x^2$ when $\Delta t / \tau \gg 1$. }
  \label{fig:0}
\end{figure}

In this case, 
the basic equations can be written as follows. 
\begin{align}
&\partial_t Q + \partial_x F = 0
\label{eq:eq5}
,
\\
&F = - \eta \partial_x Q + (F_0 + \eta \partial_x Q ) e^{- \frac{\Delta t}{\tau}}
\label{eq:eq6}
.
\end{align}
Since we use PES, 
the relaxation equation of $F$ is replaced by the formal solution. 
Then, 
we differentiate Eqs. (\ref{eq:eq5}) and (\ref{eq:eq6}). 
In the case of the second-order accuracy in time, 
\begin{align}
&\frac{Q^{n+1}_j - Q^n_j}{\Delta t} + \frac{F^{n+1/2}_{j+1/2} - F^{n+1/2}_{j-1/2}}{\Delta x} = 0
,
\label{eq:dif_adv}
\\
&F^{n+1}_j = - \eta \frac{Q^{n+1/2}_{j+1/2} - Q^{n+1/2}_{j-1/2}}{\Delta x} 
+ \left[F^n_j + \eta \frac{Q^{n+1/2}_{j+1/2} - Q^{n+1/2}_{j-1/2}}{\Delta x} \right] e^{- \frac{\Delta t}{\tau}}
.
\label{eq:dif_dif}
\end{align}
We substitute into the above equations the following forms of $Q^n_j$ and $F^n_j$ 
\begin{equation}
  Q^n_j = R^n e^{ij \theta}, \quad F^n_j = G^n e^{i j \theta}
  ,
\end{equation}
where $R^n$ and $G^n$ are the n-th power of the constants $R$ and $G$. 
Then, 
they reduce to 
\begin{align}
&R^{n+1} - R^n + 2 \frac{\Delta t}{\Delta x} i \sin \frac{\theta}{2} G^{n+1/2} = 0
\label{eq:eq7}
,
\\
&G^{n+1} = - \frac{2 \eta}{\Delta x} i \sin \frac{\theta}{2} R^{n+1/2} 
+ G^n e^{- \frac{\Delta t}{\tau}} 
+ \frac{2 \eta}{\Delta x} i \sin \frac{\theta}{2} R^{n+1/2} e^{- \frac{\Delta t}{\tau}}
\label{eq:eq8}
.
\end{align}
From these equations, 
we can obtain the following forms of equation 
\begin{equation}
R^{n+1} - \left[ 1 + e^{- \frac{\Delta t}{\tau}} 
- 4 \frac{\Delta t}{\Delta x^2} \eta (1 - e^{- \frac{\Delta t}{\tau}}) \sin^2 \frac{\theta}{2} 
 \right] R^n 
+ e^{ - \frac{\Delta t}{\tau}} R^{n-1} = 0
. 
\end{equation}
By solving this equation, 
we obtain 
\begin{equation}
R = \frac{1}{2} 
\left[
1 + e^{- \frac{\Delta t}{\tau}} 
- 4 \frac{\Delta t}{\Delta x^2} \eta (1 - e^{- \frac{\Delta t}{\tau}}) \sin^2 \frac{\theta}{2} 
\right]
\pm \sqrt{ \frac{1}{4} 
\left[
1 + e^{- \frac{\Delta t}{\tau}}
- 4 \frac{\Delta t}{\Delta x^2} \eta \sin^2 \frac{\theta}{2} 
(1 - e^{- \frac{\Delta t}{\tau}})
\right]^2
- e^{- \frac{\Delta t}{\tau}}
}
.
\label{eq:vonNeumann_PES}
\end{equation}
For the stability, 
it is necessary to satisfy the criterion $|R| \le 1$. 
In order to obtain the stability restriction, 
we substitute $|R| = 1$ for Eq. (\ref{eq:vonNeumann_PES}), 
and the equation reduces to 
\begin{equation}
  \Delta t_{min} = 
  \begin{cases}
    \frac{\Delta x^2}{2 \eta \sin^2 \frac{\theta}{2}}  & \textrm{if} \quad \Delta t \gg \tau \\
    \frac{\Delta x}{\sin \frac{\theta}{2}} \sqrt{\frac{\tau}{\eta}} & \textrm{if} \quad \Delta t \ll \tau
  \end{cases}
  \label{eq:stable1}  
  .
\end{equation}
The exact relation is given in Fig. \ref{fig:0}. 
Eq. (\ref{eq:stable1}) and Fig. \ref{fig:0} show that 
the stability restriction of the upper limit for $\Delta t$ for our numerical scheme with PES 
is the same dependence on $\Delta x$ as that of the parabolic equation 
$\Delta t < \Delta x^2 / \eta$ 
when $\Delta t \ge \tau$, 
and it becomes the same dependence on $\Delta x$ as that of the hyperbolic equation 
$\Delta t < \Delta x \sqrt{\tau / \eta}$ 
when $\Delta t \le \tau$. 
\footnote{
In addition, 
when one adopts first-order accuracy in time, 
the stability restriction becomes $\Delta t < \Delta x^2 / 4 \eta$ when $\Delta t > \tau$. }

Next, 
we solve the telegrapher equation Eq. (\ref{eq:teleg}) numerically 
by using PES method for the relaxation equation Eq. (\ref{eq:relax_F}), 
and study the dependence on the parameters. 
We take the number of grid points $N = 400$, 
and use the CFL number $0.4$. 
For the stability restriction, 
we use Eq. (\ref{eq:stable1}). 
The initial left and right states are given by 
\begin{align}
Q^L =& 3.0 & \mathrm{for} \quad & x < 1 
,
\\
Q^R =& 1.0 & \mathrm{for} \quad & x \ge 1 
.
\end{align}
We set the initial value of $F$ as $F = 0$. 

\begin{figure}[here]
 \centering
  \includegraphics[width=7cm]{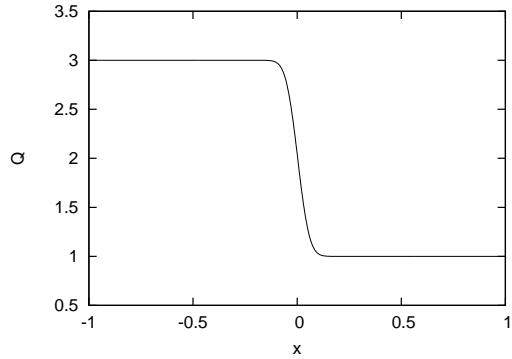}
  \caption{Numerical solution of telegrapher equation Eq. (\ref{eq:teleg}). 
           A snapshot at $t = 1.0$ is shown. 
           Diffusion smooths the initial discontinuity .
           For this test problem, 
           a cell size $\Delta x = 0.005$ is used, 
           and we set $dt = 5.0 \times 10^{-3}$, $\tau = 2.0 \times 10^{-3}$, and $\eta = 1.0 \times 10^{-3}$.}
  \label{fig:1}
\end{figure}

First, 
we set the relaxation timescale $\tau=2.0 \times 10^{-3}$, 
and the dissipation coefficient $\eta = 1.0 \times 10^{-3}$. 
Fig. \ref{fig:1} is the numerical result of PES method, 
and it reproduces the diffusion of the $Q$ very well. 

\begin{figure}[here]
 \centering
  \includegraphics[width=7cm]{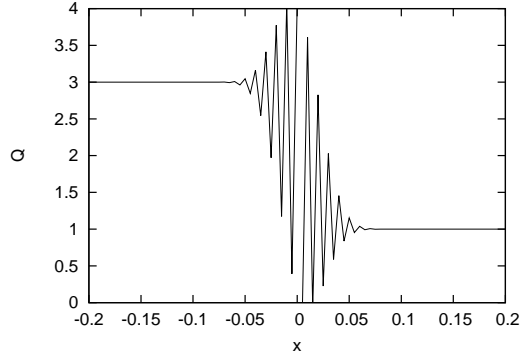}
  \caption{Numerical solution of telegrapher equation Eq. (\ref{eq:teleg}) 
           in the case of violating the stability condition ($\Delta t > \Delta t_{min}$). 
           A snapshot at $t = 0.25$ is shown. 
           The unstable modes grow quickly. 
           For this test problem, 
           a cell size $\Delta x = 0.005$ is used, 
           and we set $dt = 1.375 \times 10^{-2}$, $\tau = 2.0 \times 10^{-3}$, and $\eta = 1.0 \times 10^{-3}$.}
  \label{fig:3}
\end{figure}

Next, 
we consider the case of violating the stability restriction Eq. (\ref{eq:stable1}). 
In this case, 
we set the relaxation timescale $\tau = 2.0 \times 10^{-3}$, 
the dissipation coefficient $\eta = 1.0 \times 10^{-3}$, 
and set the timestep $\Delta t = 1.1 \Delta x^2 / 2 \eta$. 
Fig. \ref{fig:3} is the numerical result, 
and it shows that the solution diverges 
when one violates the stability restriction obtained by the von Neumann's stability analysis. 

\subsection{\label{sec:sec3.2}Stability conditions of new numerical scheme for Israel-Stewart theory}
The results of previous section will be able to be applied to the new numerical scheme for the Israel-Stewart theory, 
since the structure of the equations of the Israel-Stewart theory are telegrapher equations. 
For this reason, 
we impose the following stability conditions as the CFL condition
\begin{equation}
  \Delta t = C_a
  \begin{cases}
  \mathrm{min} 
  \left\{ \Delta x /c_s, \alpha \rho h \Delta x^2 / \mathrm{max\{ \kappa, \eta, \zeta \}}  \right\} 
  & \textrm{if} \quad \Delta t > \tau 
  ,
  \\
  \mathrm{min} 
  \left\{ \Delta x / c_s, \Delta x \sqrt{\tau / \eta} \right\} 
  & \textrm{if} \quad \Delta t < \tau
  \end{cases}
\end{equation}
where $c_s$ is the sound velocity, 
$\alpha$ is $1 / 2$ when second-order accuracy in time 
and $1 / 4$ when first-order accuracy in time, 
and $0 < C_a < 1$ is the Courant number. 
The above stability conditions are just provisional ones, 
since the dissipative RHD equations are highly non-linear equations, 
and it is difficult to obtain the exact conditions. 
Practically, the above conditions work well in our calculation. 

Next, 
%
when one solves the Israel-Stewart theory by using conservation-law form, 
one has to recover primitive variables from conservative ones. 
Then, 
if one considers large dissipation coefficients, 
the dissipation part of conservative variables $U_{dissip}$ can be in general equal or larger than $U_{ideal}$. 
This makes the primitive recovery unstable, 
since the error of dissipation variables affects considerably conserved variables $U$ 
similar to the low $\beta$ case of MHD equation. 
This may happen 
when the previous condition $\Delta x < \sqrt{\tau \eta}$ is violated. 
This is because 
$\sqrt{\tau \eta}$ is equivalent to the mean free path in ideal gas, 
and $\Delta x < \sqrt{\tau \eta}$ means 
that one resolves length scale shorter than the mean free path. 
Since the Israel-Stewart theory is approximation of the Boltzmann equation, 
the approximation becomes bad in this region. 
For this reason, 
one cannot use Israel-Stewart theory in such parameters. 
In contrast, 
if one considers effective theory, 
we recommend to impose the following restrictions on the dissipation coefficients 
\begin{align}
  q^i &< \rho h \gamma_{cs} c_s
  ,
  \\
  \tau^{ij} &< \rho h \gamma_{cs}^2 c_s^2
  ,
\end{align}
where $c_s$ is the sound velocity, 
and $\gamma_{cs} \equiv 1 / \sqrt{1 - c_s^2}$. 
We recommend to impose this restriction 
during the evolution step of dissipation variables, 
the calculation of the numerical flux, 
and the primitive recovery procedure. 
Similar conditions can be found in the previous studies~\cite{Molnar et al.(2010)}. 

In addition, 
if one adopts $\tau^{0x}, \tau^{0y}, \tau^{0z}, \tau^{yz}, \tau^{zx}, \tau^{xy}$ 
for the primitive variables, 
one needs $\tau^{xx}$ for the numerical flux. 
In this case, 
if one calculates $\tau^{xx}$ by the orthogonality condition Eq. (\ref{eq:orth1}), 
the numerical calculation sometimes becomes unstable, 
since the calculation includes the division by the velocity. 
In order to prevent this numerical divergence, 
we use the Navier-Stokes term for $\tau^{xx}$. 
This approximation becomes bad 
when the timestep $\Delta t$ is close to the relaxation time. 
However, 
if one considers the region where the fluid approximation is not so bad valid, 
the relaxation time is small 
and the above approximation gives accurate value for $\tau^{xx}$. 

\section{\label{sec:sec4}Test Calculations}
In this section, 
results of several one-dimensional and multi-dimensional test simulations are presented. 
In the following test problems, 
we use CFL number $0.4$, 
and we consider an ideal equation of state $h \equiv 1 + \Gamma / (\Gamma - 1) p / \rho$. 
\subsection{\label{sec:sec4.1}1D test}
\subsubsection{\label{sec:sec4.1.1}shear flow}
For the stability and causality, 
the Israel-Stewart theory adds $9$ variables to the $5$ fluid variables, 
and it is very difficult to obtain exact solutions. 
However, 
relaxation of tangential velocity by the shear viscosity can be expected 
that the solution approaches to the exact solution of the non-relativistic case 
if one uses sufficiently low velocity. 
Thus, 
we consider the following initial condition 
\begin{align}
(\rho^L, p^L, v^{yL}) =& (1.0, 1.0, - 0.1) & \mathrm{for} \quad x < 0.0
,
\\
(\rho^R, p^R, v^{yR}) =& (1.0, 1.0,   0.1) & \mathrm{for} \quad x \ge 0.0
.
\end{align}
All the other fields are set to $0$. 

\begin{figure}[here]
 \centering
  \includegraphics[width=7cm]{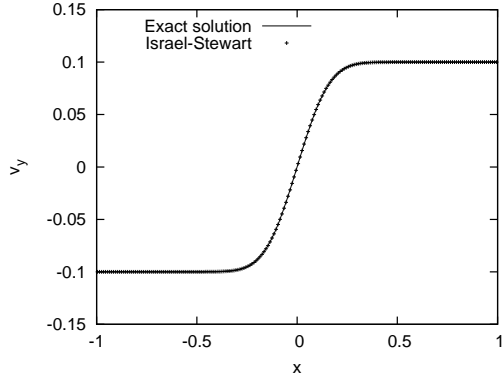}
  \caption{Numerical solution of the relaxation of a shear flow. 
           Crosses denote a snapshot at $t = 4.0$. 
           The analytical solution Eq. (\ref{eq:vy_exact}) is shown as a solid curve. 
           Our scheme for Israel-Stewart theory reproduces the analytical solution very well. 
           For this test problem, 
           a cell size $\Delta x = 0.01$ is used, 
           and we set $\Gamma = 4 / 3$, and $\eta = 0.01$.}
  \label{fig:test1}
\end{figure}

In this case, 
the relativistic Euler equation is given by 
\begin{equation}
\rho h \partial_t [\gamma^2 v^y] + \partial_x \tau^{xy} = 0
\label{eq:EoM_shear}
.
\end{equation}
Since we consider sufficiently low velocity, 
the Lorentz factor $\gamma$ is nearly unity. 
The viscous tensor $\tau^{\mu \nu}$ reduces to 
\begin{equation}
\tau^{xy} = - \eta \partial^x v^y
\label{eq:test_shear}
.
\end{equation}
Using Eqs. (\ref{eq:EoM_shear}) and (\ref{eq:test_shear}), 
the following exact solution can be obtained 
\begin{align}
v^y &= v^y_0 \mathrm{erf} \left[ \frac{|x - x_0|}{\chi t} \right]
\label{eq:vy_exact}
,
\\
\chi &= \frac{\eta}{\rho h}
.
\end{align}
Fig. \ref{fig:test1} is the numerical results at $t = 4.0$. 
We use an ideal equation of state with $\Gamma = 4 / 3$, 
a cell size $\Delta x = 0.01$ is used, 
and the viscosity coefficient $\eta = 0.01$. 
Fig. \ref{fig:test1} shows that 
the numerical solution reproduces the exact solution very well. 
The convergence test shows that 
it is consistent with the second-order accuracy. 

\subsubsection{\label{sec:sec4.1.2}shock tube test}
In this section, 
we consider the shock tube problem. 
When one describes hydrodynamics including shock waves by using ``ideal'' fluid solver, 
the thickness of the discontinuities are determined by the numerical dissipation. 
In reality, 
the thickness of the shock wave front is determined by dissipation coefficients. 
In the following, 
we show that 
the thickness of the discontinuities depends on the dissipation coefficients of our code. 
The exact Riemann Solver for ideal fluid part is based on 
the solution by Mart{{\'\i}} and M$\ddot{\mathrm{u}}$ller~\cite{Marti & Muller(2003)}. 

We prescribe the initial left and right states 
\begin{align}
(\rho^L, p^L, v^{yL}) =& (10.0, 10.0, 0.2) & \mathrm{for} \quad & x < 0.5
,
\\
(\rho^R, p^R, v^{yR}) =& (1.0, 1.0, - 0.2) & \mathrm{for} \quad & x \ge 0.5
,
\end{align}
with an ideal equation of state with $\Gamma = 5 / 3$, 
and all the other fields are set to $0$. 
Integration is carried until $t = 0.4$, 
and a cell size $\Delta x = 0.0025$ is used. 
The dissipation coefficients 
thermal conductivity $\kappa$, shear viscosity $\eta$, and bulk viscosity $\zeta$ 
are assumed constant, 
and set $10^{-15}$ 
unless stated otherwise. 

\begin{figure}[here]
 \centering
  \includegraphics[width=7cm]{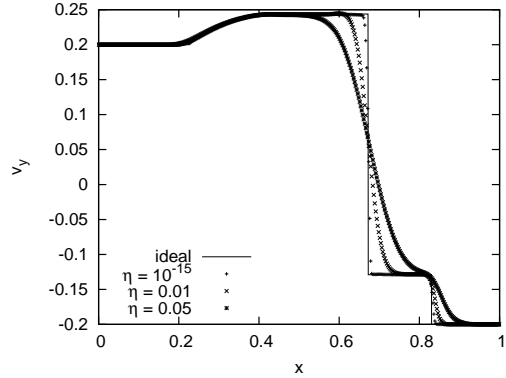}
  \caption{The numerical solution for the relativistic shock tube problem. 
           Tangential velocity profile at $t = 0.4$ are shown by crosses 
           in the cases of three different shear viscosity coefficients:$\eta = 10^{-15}, 0.01, 0.05$. 
           The result of ideal gas is also shown by a solid line. 
           In this test problem, 
           the widths of discontinuities are determined by the \textit{physical} viscosity. 
           For this test problem, 
           a cell size $\Delta x = 0.0025$ is used, 
           and we set $\Gamma = 5 / 3$, and the CFL number $0.4$.}
  \label{fig:shear}
\end{figure}

First, 
Fig. \ref{fig:shear} is the numerical results of the tangential velocity $v^y$ that 
changes the shear viscosity $\eta = 0.01, 0.05, 10^{-15}$ 
compared to the RHD exact solution. 
It shows that 
both contact discontinuity and shock are diffused, 
and the width of the discontinuity of $\eta = 0.05$ is about $5$ times greater than $\eta = 0.01$.

\begin{figure}[here]
 \centering
  \includegraphics[width=7cm]{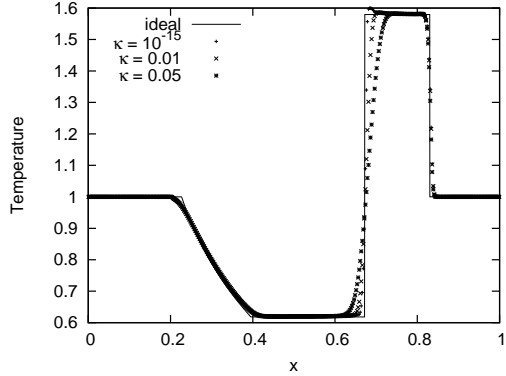}
  \caption{The numerical solution for the relativistic shock tube problem. 
           Temperature profiles at $t = 0.4$ are shown by crosses 
           in the cases of three different thermal conduction coefficients:$\kappa = 10^{-15}, 0.01, 0.05$. 
           All the profiles are similar near the shock wave, 
           but the width of contact discontinuity is determined by the thermal conductivity. 
           For this test problem, 
           a cell size $\Delta x = 0.0025$ is used, 
           and we set $\Gamma = 5 / 3$, and the CFL number $0.4$.}
  \label{fig:therm}
\end{figure}

Next, 
Fig. \ref{fig:therm} is the numerical results of the temperature $T \equiv p / \rho$ 
for various values of the thermal conductivity $\kappa = 0.01, 0.05, 10^{-15}$ 
compared to the RHD exact solution. 
It shows that 
the heat flows from high temperature region to low temperature region, 
and both contact discontinuity and shock are smoothed, 
and the width of the discontinuity of $\kappa = 0.05$ is about $5$ times greater than $\kappa = 0.01$. 

\begin{figure}[here]
 \centering
  \includegraphics[width=7cm]{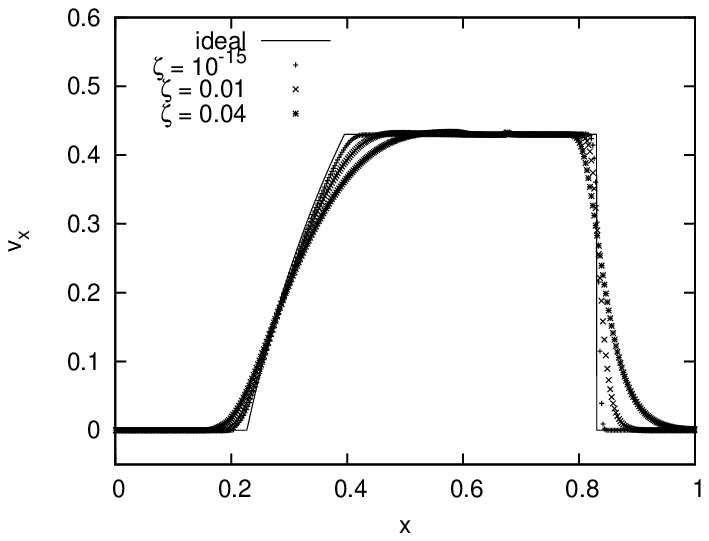}
  \caption{The numerical solution for the relativistic shock tube problem. 
           The profile of longitudinal velocity $v^x$ at $t = 0.4$ is shown 
           for three different shear viscosity coefficients:$\zeta = 10^{-15}, 0.01, 0.05$. 
           The widths of discontinuities and rarefaction fronts are determined by the \textit{physical} viscosity. 
           For this test problem, 
           a cell size $\Delta x = 0.0025$ is used, 
           and we set $\Gamma = 5 / 3$, and the CFL number $0.4$.}
  \label{fig:bulk}
\end{figure}

Fig. \ref{fig:bulk} is the numerical results of the velocity $v^x$ 
for various value of the bulk viscosity $\zeta = 0.01, 0.04, 10^{-15}$ 
compared to the RHD exact solution. 
The bulk viscosity diffuses the fluid expansion $\theta = u^{\mu}_{\mu}$, 
and in the one-dimensional case 
the expansion reduces to $\theta = \partial_x u^x$. 
Fig. \ref{fig:bulk} shows that 
the gradient of $v^x$ in x-direction is smoothed out. 

\subsection{\label{sec:sec4.2}Two-dimensional Kelvin-Helmholtz Instability}
In this section, 
we present the numerical results of the two-dimensional Kelvin-Helmholtz instability (KH instability) 
for the multidimensional numerical test problem. 
Multidimensional extension can be achieved simply via directional splitting 
explained in Sec.~\ref{sec:sec2.4}. 

KH instability is that of a tangential discontinuity between parallel flows. 
It is well-known from the linear perturbation analyses that 
this instability arises when the Lorentz factor of fluid is not so high, 
and the growth rate is proportional to the wave number. 
Second property means that 
the numerical simulation of KH instability does not converge in the ideal fluid case; 
if one increases the number of grid points, 
grid-size-scale perturbations grow most rapidly, 
and this prevents numerical convergence of KH instability in the ideal fluid case. 
However, 
if one considers dissipation, 
the numerical convergence is possible 
since the perturbations shorter than the characteristic scale of dissipation are smoothed out. 

The initial condition is prescribed as 
\begin{align}
(\rho, p, v^x, v^y) =& (1, 0.3, 0.1, 0.0) & \mathrm{for} \quad & y > 0.0
,
\\
(\rho, p, v^x, v^y) =& (2, 0.3, - 0.1, 0.0) & \mathrm{for} \quad & y \le 0.0
.
\end{align}
To trigger the KH instability, 
we perturb the shear flow by the position of tangential discontinuity $y_{tangential}$
\begin{align}
  y_{tangential} &= 0.01 \sin(k_x x)
  ,
  \\
  k_x &= 2 \pi
\end{align}
We use an equation of state with $\Gamma = 5 / 3$, 
and the non-relativistic limit of relaxation time Eq. (\ref{eq:relax_nonrela}). 
The computational domain covers the region $[-1,1] \times [-1,1]$ with $1024 \times 1024$ grid points. 
The CFL number is $0.2$, 
and the integration is carried out until $t = 30$, 
that is, about $1.5$ fluid crossing time. 
We set periodic boundary condition for x-direction, 
and reflecting boundary condition for y-direction for simplicity. 

\begin{figure}[here]
 \centering
  \includegraphics[width=8cm]{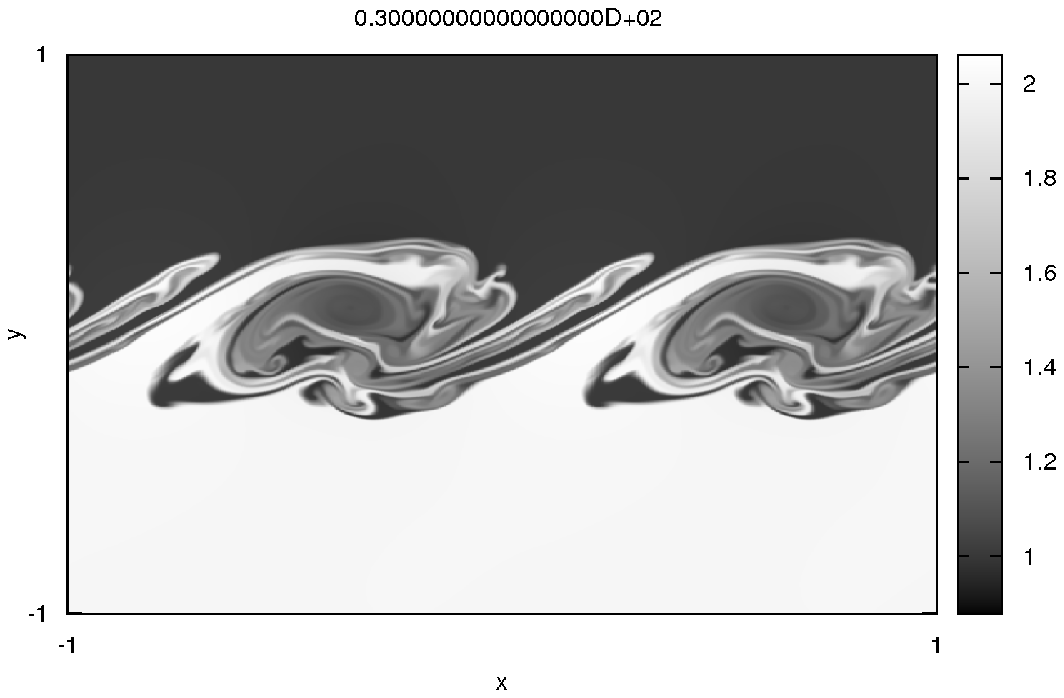}
  \caption{Non-linear development of Kelvin-Helmholtz instability in two-dimensional simulations without viscosity. 
           The density profile at $t = 30$ is presented. 
           Numerical integration has been performed with $\Delta t = 0.2 \Delta x / c_s$. 
           Rolling up of the interface (tangential discontinuity) is the non-linear consequence 
           of the rapid growth of the initial perturbation of ``sine'' function, 
           characteristics of the KH instability. 
           However, 
           numerical cell-size-scale perturbations also grow, 
           and contaminate the result. 
           }
  \label{fig:KH1}
\end{figure}

\begin{figure}[here]
 \centering
  \includegraphics[width=8cm]{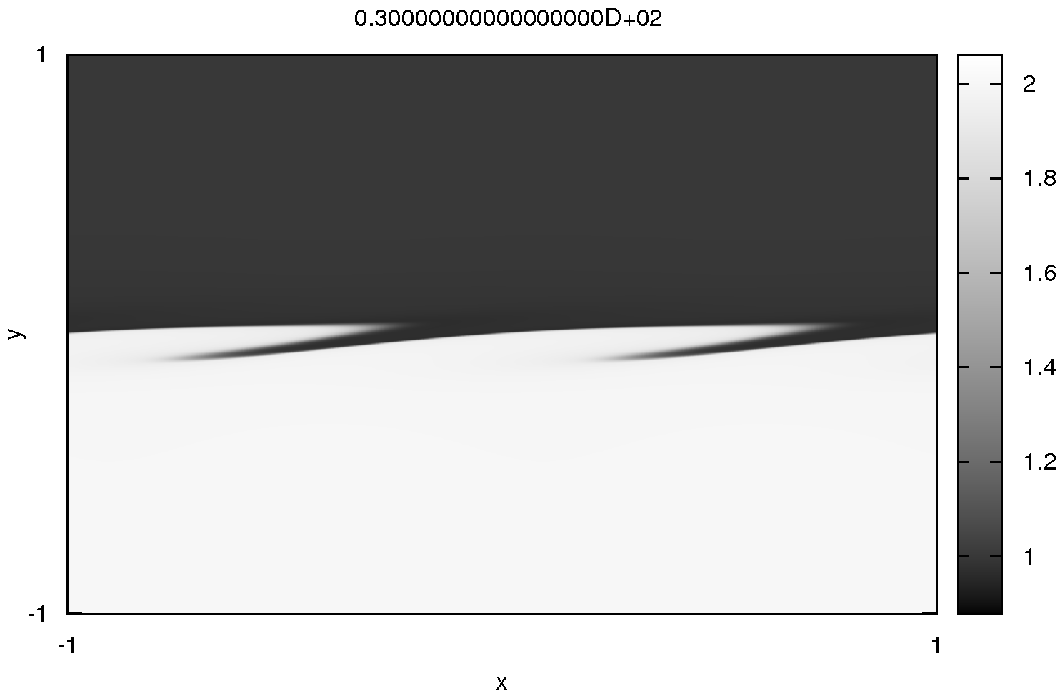}
  \includegraphics[width=8cm]{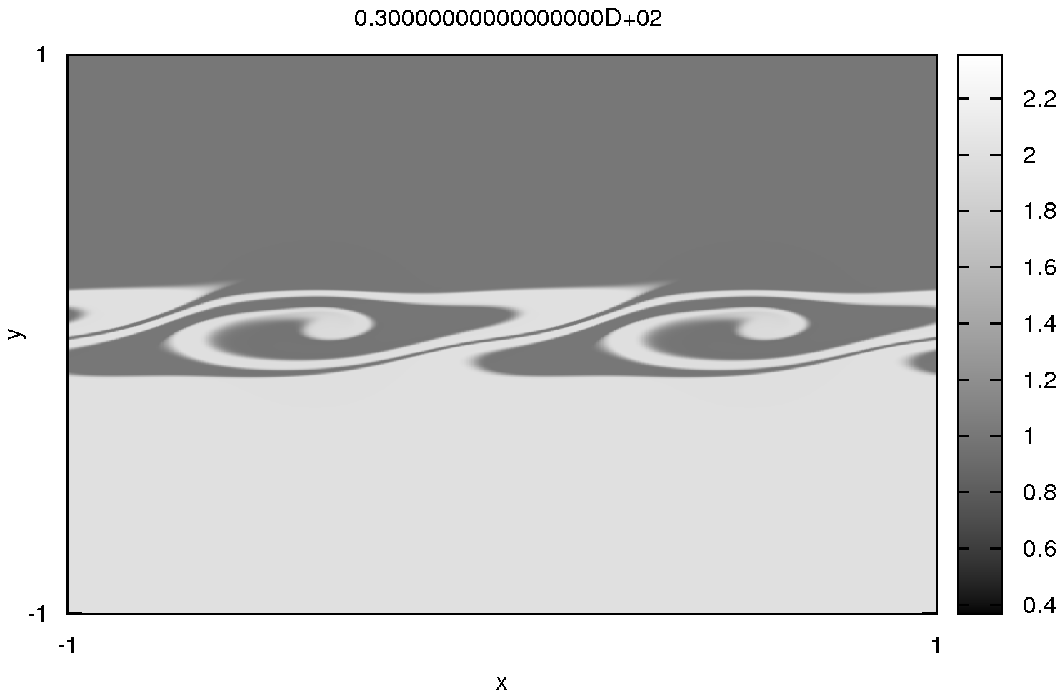}
  \caption{Non-linear development of Kelvin-Helmholtz instability in two-dimensional simulation with viscosity. 
           The density profile at $t = 30$ is presented. 
           Numerical integration has been performed with $\Delta t = 0.2 \Delta t_{min}$, 
           where $\Delta t_{min}$ is presented in Eq. (\ref{eq:stable1}). 
           We set the shear viscosity $\eta = 5 \times 10^{-3}$ for the upper fluid, 
           and $\eta = 10^{-3}$ for the lower fluid. 
           Numerical cell-size-scale perturbations are stabilized by the viscosity. }
  \label{fig:KH2}
\end{figure}

Fig. \ref{fig:KH1} is the result of ideal fluid case 
carried by using a relativistic Godunov scheme~\cite{Mignone et al.(2005)}
for the sake of comparison. 
This figure shows that 
the rolling up of the interface results from the KH instability. 
Note that 
numerical grid-size-scale perturbations grow 
in addition to the initial perturbation of $\sin$ function mode. 
Figs. \ref{fig:KH2} are the result of our new code of Israel-Stewart theory. 
In this simulation, 
we consider only shear viscosity, 
and use $\eta = 0.005$ and $\eta = 0.001$ as the dissipation coefficient. 
Figs. \ref{fig:KH2} show that 
numerical grid-size-scale perturbations are stabilized, 
and only the initial perturbation of sine mode forms vortices. 
Note that 
for avoiding the numerical grid-size-scale perturbation in inviscid case, 
one has to introduce a sufficiently large scale of gradient $\alpha$ to the profile of fluid variables $d, v_x$ as follows: 
\begin{equation}
  Q = \bar{Q} + \Delta Q \tanh(y / \alpha)
  ,
\end{equation}
where $\bar{Q} = (Q_{y>0} + Q_{y<0}) / 2$ and $\Delta Q = |Q_{y>0} - Q_{y<0}|$. 

\begin{figure}[here]
 \centering
  \includegraphics[width=8cm]{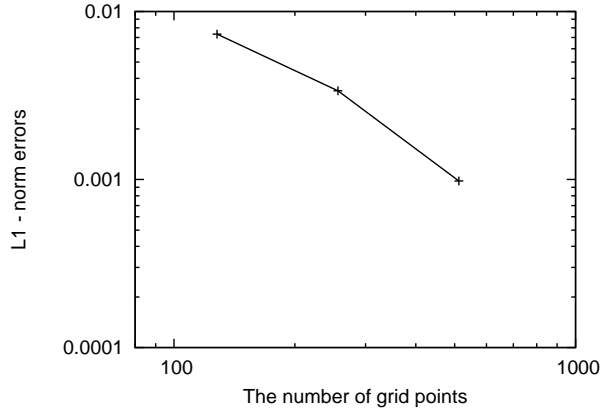}
  \caption{$L_1$ norm errors of the density of the 2-dimensional shear flow 
           calculated by the new dissipation code 
           in the case of $\eta = 0.001$. 
           This figure shows that 
           the numerical solutions converge to the result of $1024$ grid points 
           with the second-order accuracy. 
           }
  \label{fig:KH3}
\end{figure}

Fig. \ref{fig:KH3} is $L_1$ norm errors of the density under different grid points 
carried out by the new dissipation code 
in the case of $\eta = 0.005$. 
The solutions are compared to the result of $1024$ grid points. 
This figure shows that 
the numerical results converge to the result of $1024$ grid points 
with the second-order accuracy 
because of the shear viscosity. 

\begin{figure}[here]
 \centering
  \includegraphics[width=8cm]{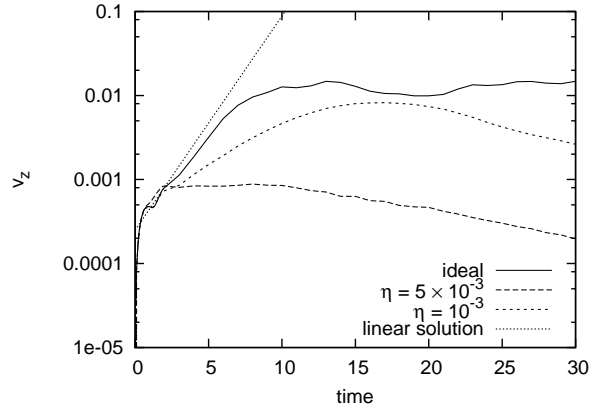}
  \caption{Evolution of the amplitude of $v_y$ as a function of time 
           comparing to the analytical solution of the linear perturbation 
           obtained by Turland and Scheuer~\cite{Turland & Scheuer(1976)}. 
           The numerical integration is calculated by the new dissipation code 
           in the case of $\eta = 0.005, 0.001$, and inviscid case. 
           This figure shows that 
           the numerical solutions reproduce the analytical theory in the linear growth region. 
           }
  \label{fig:KH4}
\end{figure}

In Fig. \ref{fig:KH4}, 
we compare the time evolution of the amplitude of $\bar{v_y}$ 
that is a square root of the sum total of $v_y^2$ 
to the analytical solution of the linear perturbation 
obtained by Turland and Scheuer~\cite{Turland & Scheuer(1976)}. 
The numerical integration is carried out by the new dissipation code 
in the case of $\eta = 0.005, 0.001$, and inviscid case. 
During the initial phase, 
the evolution of $v_y$ reproduces the prediction of the linear theory well. 
After the linear phase, 
the numerical solutions grow non-linearly, 
and start to deviate from linear theory 
because of the effect of shear viscosity and numerical grid-size-scale perturbation. 
Note that 
when the shear viscosity $\eta$ takes larger value, 
the growth of $v_y$ saturates faster, 
and decay in time. 


\section{\label{sec:sec5}Conclusion}
In this paper, 
we have presented a new numerical scheme for relativistic dissipative hydrodynamics, 
that is, Israel-Stewart theory. 
Israel-Stewart theory is a stable and causal relativistic dissipation theory. 
However, for the stability and causality, 
this theory includes relaxation equations of dissipation variables. 
In general, 
relaxation timescales of dissipation variables are much shorter than characteristic timescale of hydrodynamics. 
This means that 
relaxation equations of dissipation variables are stiff equations, 
and this makes it difficult to integrate Israel-Stewart theory numerically. 
In our new scheme, 
we use Strang's splitting method, 
and obtain formal solution of the relaxation equation 
for solving this extremely short timescale problem. 
By using the formal solution, 
the Courant condition of relaxation time disappear, 
since we do not use explicit finite difference scheme. 
In addition, 
since we split the calculation of inviscid step and dissipation step, 
Riemann solver can be used for obtaining numerical flux of inviscid part, 
and this enables us to obtain more accurate solution. 

In astrophysical application, 
it is very important to take into account dissipation terms, 
since dissipation terms transform kinetic energy of bulk fluid into thermal energy, 
which becomes observable as thermal radiation. 
In addition, 
if one considers the dynamics of accretion disk, 
$\alpha$ viscosity is often used for the phenomenological models 
of angular momentum transfer, 
and our new scheme can be used for those modeling. 
In recent years, 
the strongly coupled quark-gluon plasma (QGP) 
in the Relativistic Heavy-Ion Collider (RHIC) 
has been vigorously studied 
by using description by relativistic dissipative hydrodynamics equations 
in nuclear physics, 
and our scheme will be useful for such calculations. 

\section*{acknowledgments}
Numerical computations were [in part] carried out on Cray XT4 at Center for Computational Astrophysics, 
CfCA, of National Astronomical Observatory of Japan.


\appendix

\end{document}